\documentclass[journal]{IEEEtran}

\usepackage{xcolor,soul,framed} 

\colorlet{shadecolor}{yellow}
\usepackage[pdftex]{graphicx}
\graphicspath{{../pdf/}{../jpeg/}}
\DeclareGraphicsExtensions{.pdf,.jpeg,.png}

\usepackage[cmex10]{amsmath}
\usepackage{array}
\usepackage{mdwmath}
\usepackage{cancel}
\usepackage{mdwtab}
\usepackage{eqparbox}
\usepackage{url}
\usepackage{amsfonts}
\usepackage{booktabs}
\usepackage{amsthm}
\usepackage{multicol}
\usepackage{listings}
\usepackage{graphicx}

\lstdefinelanguage{Cisa}{%
  language     = C,
  morekeywords = {asm, posit_t, __asm__, . set, . byte},
}

\lstdefinelanguage{Cisa2}{%
  language     = C,
  morekeywords = {padd, pmul, gemm, posit_t},
}

\newcommand{\colorbitbox}[3]{%
\rlap{\bitbox{#2}{\color{#1}\rule{\width}{\height}}}%
\bitbox{#2}{#3}}

\definecolor{lightcyan}{rgb}{0.84,1,1}
\definecolor{lightgreen}{rgb}{0.64,1,0.71}
\definecolor{darkgreen}{rgb}{0,0.6,0}
\definecolor{lightorange}{rgb}{1,0.9,0.8}
\definecolor{lightred}{rgb}{1,0.7,0.71}
\definecolor{lightcyan}{rgb}{0.84,1,1}
\definecolor{lightgreen}{rgb}{0.64,1,0.71}
\definecolor{lightblue}{rgb}{0.8,0.8,1}
\definecolor{amber}{rgb}{1.0, 0.75, 0.0}
\definecolor{darkamber}{rgb}{0.8, 0.65, 0.0}
\definecolor{lightmagenta}{rgb}{1,.84,1}
\definecolor{brightblue}{rgb}{0.25,0.5,1.}

\usepackage{algorithm}
\usepackage{algpseudocode}
\usepackage{amssymb}
\usepackage{multirow}
\usepackage{bytefield}
\usepackage[many]{tcolorbox}

\usepackage{cite}
\begin{document}
\title{FPPU: Design and Implementation of a Pipelined Full Posit Processing Unit}
 \author{Federico~Rossi$^1$,
      Francesco~Urbani$^1$,
      Marco~Cococcioni$^1$,
      Emanuele~Ruffaldi$^2$,
      Sergio~Saponara$^1$

 \thanks{Work partially supported by H2020 projects EPI-SGA2 (grant no. 101036168, \protect\url{https://www.european-processor-initiative.eu/} ) and TEXTAROSSA (grant no. 956831,  \protect\url{https://textarossa.eu/}  ).}
 \thanks{$^1$Federico Rossi, Francesco Urbani, Marco Cococcioni, Sergio Saponara are with University of Pisa, Department of Information Engineering, Pisa, Italy (e-mail: federico.rossi@ing.unipi.it, [name.surname]@unipi.it).}

 \thanks{$^2$Emanuele Ruffaldi is with Medical Microinstruments Inc., Italy (e-mail: emanuele.ruffaldi@mmimicro.com).}
 }

\markboth{IEEE Transactions on Emerging Topics in Computing
}{Rossi \MakeLowercase{\textit{et al.}}: FPPU: Design and Implementation of a Pipelined Full Posit Processing Unit}

\maketitle

\begin{abstract}
By exploiting the modular RISC-V ISA this paper presents the customization of instruction set with posit\textsuperscript{\texttrademark} arithmetic instructions to provide improved numerical accuracy, well-defined behavior and increased range of representable numbers while keeping the flexibility and benefits of open-source ISA, like no licensing and royalty fee and community development.
In this work we present the design, implementation and integration into the low-power Ibex RISC-V core of a full posit processing unit capable to directly implement in hardware the four arithmetic operations (add, sub, mul, div and fma), the inversion,  the float-to-posit  and posit-to-float conversions. We evaluate speed, power and area of this unit (that we have called Full Posit Processing Unit). The FPPU has been prototyped on Alveo and Kintex FPGAs, and its impact on the metrics of the full-RISC-V core have been evaluated, showing that we can provide real number processing capabilities to the mentioned core with an increase in area limited to 7\% for 8-bit posits and to 15\% for 16-bit posits. Finally we  present tests one the use of posits for deep neural networks with different network models and datasets, showing minimal drop in accuracy when using 16-bit posits instead of 32-bit IEEE floats.

\end{abstract}


%
\IEEEpeerreviewmaketitle


\section{Introduction}

\IEEEPARstart{T}{he} RISC-V (Reduced Instruction Set Computer - V) open instruction set architecture (ISA) \cite{riscvisa,riscvabout} is a modern, open-source ISA that is gaining popularity in recent years. RISC-V was designed to be a free and open alternative to proprietary ISAs such as ARM and x86, with the goal of fostering innovation and research in the computer architecture community. Unlike proprietary ISAs, RISC-V is not tied to any specific vendor or implementation and can be used in a wide range of devices, from microcontrollers to supercomputers. 

Posit\textsuperscript{\texttrademark} numbers are a recent alternative to traditional floating-point arithmetic (IEEE 754 floating point numbers: binary32, binary64, ...). The posit number system, introduced in 2017 by J. L. Gustafson \cite{gustafson2017beating} and standardized in 2022 \cite{posit_standard_2022}, is designed to provide improved numerical accuracy and range, thus maintaining a similar level of performance as traditional floating-point numbers but for reduced data bit-width. Posit numbers are represented by a fixed number of bits, just like floating-point numbers, but they use a different encoding scheme that allows for a larger range of representable numbers. Posit numbers also provide a well-defined and predictable behavior when dealing with numbers close to zero.
Multiple works proved the capabilities of Posit number to be a drop-in replacement of binary32 numbers for Deep Neural Networks (DNNs) \cite{coco_smallreals, coco_fast_elu_smartcomp, deeppositron, carmichael2019deeppos}. In particular 8-bit posits demonstrated to be able to maintain similar accuracy in DNN tasks when compared to binary32 numbers and 16-bit posits can even outperform binary32.
In this work we advance the combination of the RISC-V open instruction set and posit arithmetic to provide an efficient way to  process real numbers in RISC-V cores already seen in \cite{ppulight}, adding this capability to the small Ibex RISC-V core \cite{schiavone_zeroriscy} without using a Floating Point Unit (FPU). With respect to the Light PPU \cite{ppulight} (which supported only float-to/from-posit conversions and it was essentially used to save storage of DNN weights), the Full Posit Processing Unit (FPPU) proposed in this work also provides hardware support to the four posit operations (sub, add, mul, div) and to the computation of reciprocals. This means that,  beside storage, also computations in posit domain become fast, being native and not simulated. The RISC-V ISA is highly modular and customizable, making it well-suited for the integration of a posit processing unit. The modular nature of RISC-V allows to include only the instructions and functional units that are needed for posit arithmetic, without the need to include instructions or functional units that are not required (e.g. IEEE FPUs and related RV32F ISA). This can result in a smaller and more efficient processing core, as already proved in \cite{PACoGen,SHARMA2023102801,9737228}. On the other hand, posit arithmetic can provide improved numerical accuracy, well-defined behavior, and increased range of representable numbers, making it well-suited for applications that require these features such as embedded systems, large data sets, and scientific computing. 

\begin{table*}[t]
\centering
\caption{Similarities and differences between this work and related posit processing units}
\label{tab:related_works}
\begin{tabular}{l|c|c|c|c}
\hline
\multicolumn{5}{c}{HARDWARE STACK}                                                                \\ \hline
\multicolumn{1}{l|}{}                                                        & This work & Percival \cite{9817027}  & Clarinet \cite{SHARMA2023102801}  & Pacogen \cite{8731915}  \\
\multicolumn{1}{l|}{Configurable posit size}                                 &   $\checkmark$        & $\times$ & $\checkmark$  & $\times$  \\
\multicolumn{1}{l|}{High-precision posit support (32,64 bit)}                &   $\checkmark$        & $\checkmark$ & $\checkmark$  & $\checkmark$  \\
\multicolumn{1}{l|}{Low-precision posit support (8,16 bit)}                  &   $\checkmark$        & $\times$ & $\checkmark$  & $\checkmark$  \\
\multicolumn{1}{l|}{Quire/Fused support (FMA)}                                     &   $\checkmark$        & $\checkmark$ & $\checkmark$  & $\times$  \\
\multicolumn{1}{l|}{Dynamic power monitored}                                 &   $\checkmark$ (see \cite{parma2023piccoli})        & $\times$ &   $\times$ & $\times$  \\
\multicolumn{1}{l|}{RISC-V Integration}                                      &   $\checkmark$        & $\checkmark$  & $\checkmark$ & $\times$ \\
\multicolumn{1}{l|}{Platform and soft-core independent}                                      &   $\checkmark$        & $\times$  & $\checkmark$  & $\checkmark$  \\
\multicolumn{1}{l|}{SIMD/Vector operations}                                       &   $\checkmark$        & $\times$  & $\times$ & $\times$  \\ \hline
\multicolumn{5}{c}{SOFTWARE STACK}                                                                \\ \hline
\multicolumn{1}{l|}{High-level software support (e.g. posit library)}        &  $\checkmark$         & $\times$ & $\times$  & $\times$  \\
\multicolumn{1}{l|}{Compiler independent instruction support}                &  $\checkmark$         & $\times$ & N/A  & N/A  \\
\multicolumn{1}{l|}{Integration with Deep Neural Networks (DNNs) frameworks} &   $\checkmark$        & $\times$ & $\times$ & $\times$  \\ \hline
\end{tabular}
\end{table*}

\subsection{Organization of the paper}
Section II summarise differences and similarities of this work with other related researches. Section \ref{sec:positnms} describes posit numbers and outlines main properties. Section \ref{sec:positarith} elaborates on the logical implementation of posit operations, focusing on the different aspects of division algorithms. Section \ref{sec:fppu} shows the design of the FPPU. Section \ref{sec:rvisaposit} describes the Instruction Set Architecture (ISA) extension for posit operations and the SW support for compilation. Section \ref{sec:fppuintegr} shows the design steps for the integration of the FPPU inside the low-power Ibex core. Section \ref{sec:ppuchar} characterizes the FPPU and Ibex components for area occupation, clock and power consumption. Discussions and Conclusions are given in the last section. The IP database for the posit unit is completely contained at: \url{https://github.com/federicorossifr/ppu_public}.
\section{Related works}
There exists several hardware implementation of posit processing units, each one with different peculiarities  in terms of configurable posit size, pipeline architecture, implemented operations, resource utilisation, RISC-V integration and software support. We summarise some of them, comparing with this work in Table \ref{tab:related_works}. We highlight the hardware characteristics in terms of flexibility of the posit configuration (e.g. number of bits and number of exponent bits), support for fused operations, vectorized operations and capabilities to integrate the unit inside a RISC-V soft-core. Furthermore we elaborate on the software stack, highlighting whether the unit has an high-level support in the software domain via libraries, compilers and computational libraries.

\section{Posit Numbers}\label{sec:positnms}

\begin{figure}
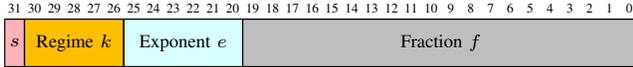

	\centering    
    \begin{bytefield}[bitwidth=0.75em]{32}
       \bitheader[endianness=big]{0-31}\\
       \small
       \colorbitbox{lightred}{1}{{\scriptsize{$s$}}}&
       \colorbitbox{amber}{5}{\scriptsize{Regime $k$}} &
       \colorbitbox{lightcyan}{6}{\scriptsize{Exponent $e$}} &
       \colorbitbox{lightgray}{20}{\scriptsize{Fraction $f$}} 
    \end{bytefield}
    \caption{Bit fields of a posit$\left<32,6\right>$  data type.}
	\label{fig:32bitPositIllustration}
\end{figure}

\begin{figure}[H]
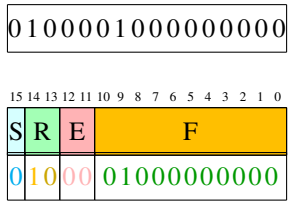
\centering\begin{bytefield}[bitwidth=0.66em]{16}\bitbox{16}{0\,1\,0\,0\,0\,0\,1\,0\,0\,0\,0\,0\,0\,0\,0\,0}\\\\\bitheader[endianness=big]{0-15}\\\colorbitbox{lightcyan}{1}{{S}}&\colorbitbox{lightgreen}{2}{R}&\colorbitbox{lightred}{2}{E}\colorbitbox{amber}{11}{F}\\\bitbox{1}{\color{cyan}0}&\bitbox{2}{\color{amber}1\!\;\color{darkamber}0}&\bitbox{2}{\color{lightred}0\!\;0}&\bitbox{11}{\color{darkgreen}0\!\;1\!\;0\!\;0\!\;0\!\;0\!\;0\!\;0\!\;0\!\;0\!\;0}&\end{bytefield}\caption{An example of Posit configuration with 16 bits and 2 exponent bits. The associated real value to the shown Posit is:$\textcolor{cyan}{+} 16^{\textcolor{darkamber}{0}}\times 2^{\textcolor{lightred}{0}}\times ( 1 + \textcolor{darkgreen}{512}/2048)= 1.25$.}\label{fig:positex}\end{figure}

A posit number (see Fig.~\ref{fig:32bitPositIllustration}) is represented by a signed integer on 2's complement. It can be configured with the total number of bits $N$ and maximum number of exponent bits $ES$. We define such a posit as $\text{Posit}\left < N,ES\right >$. The format can have at most four fields: i) sign on 1-bit, ii) regime with a variable size (run-length encoded), iii) exponent with at most $ES$ bits and iv) fraction with a variable length. An example of a posit number instance is shown in Fig.~\ref{fig:positex}. Note that, if the regime fields is large enough, it is possible that the exponent field has less bits available than $ES$. In this case the actual exponent value is computed by padding zeroes to the right of the exponent bits in the format.

The length $l$ of the regime corresponds to the number of identical bits following the sign bit:
\begin{equation}    
    s,\underbrace{b_1, b_2, b_3, \dots, b_l}_{=b},\underbrace{b_{l+1}}_{=\overline{b}}, \cdots
\end{equation}
The regime length is $l$. Depending on the value of the bit $b$, the regime value $k$ will be computed as follows:
\begin{equation}\label{eqn:regimeValue}
k := \left\{\begin{matrix}
 l-1& \, \textnormal{if} & b = 1 \;\; \\
 -l & \, \textnormal{if} & b = 0 \; . \\
\end{matrix}\right.
\end{equation}

\noindent The regime value is a scale factor for a special constant, that depends on the posit configuration, called \textit{useed}. The \textit{useed} value is computed as follows:
\begin{equation}\label{eqn:useed}
    \text{useed} := 2^{2^{ES}} 
\end{equation}

\noindent Hence, the real value $r$ associated with a posit represented by the integer $P$ on two's complement (with sign $s$) is computed as in Equation \eqref{eqn:positRealValue}: 

\begin{equation}\label{eqn:positRealValue}
 r :=   \left\{\begin{matrix}
0 & \textnormal{if}\; P = 0  \\
\textnormal{NaR (Not a Real)} & \textnormal{if}\; P = -2^{N-1} \\
(-1)^s \times \text{useed}^{k} \times 2^e \times \left ( 1+ \frac{f}{2^F} \right) &  \textnormal{otherwise} \, .\\
\end{matrix}\right.
\end{equation}

The value $F$ is the length of the fraction field. Note that there will be always an implicit one in front of the fraction (i.e. $1.f_1,f_2, \dots, f_F$), without any subnormal number differently from IEEE binary32 numbers.

In \cite{posit_standard_2022} there was reported an alternative version of the third expression of Equation \eqref{eqn:positRealValue}, which is: 
\begin{equation}\label{eqn:rtp}
r= (1-3s+f/2^F) \times 2^{(1-2s)\times (2^{\textit{ES}}\times k+e+s)} \, .
\end{equation}

\section{Posit arithmetic operations and implementation}\label{sec:positarith}

This Section elaborates on the logic implementation of posit arithmetic operations, deepening on the arithmetical correctness and numerical rounding of such operations. Each posit operation described here starts with a \textit{decoding} and \textit{input conditioning} phase, where the posit operands are decoded into the \textit{sign, regime, exponent, significand} fields and decisions are made depending on few special cases (for example when one of the operands is equal to $0$ or NaR). After this phase, each posit operand is transformed into a general Floating-point Intermediate Format (FIR) $\left <s_f,te,f \right >$, where $s_f$ is the sign, $te$ is the total exponent (without bias) and $f$ is the fractional part of the significand (\textbf{note: the FIR representation is just a zero-cost transformation of a decoded posit and do not share any similarity with the IEEE 32-bit floating point (e.g. no NaN, Infs or subnormals))}:
\begin{equation*}
        p = (-1)^{s_f} \times 2^{te} \times (1.f) \; . 
\end{equation*}
While $s_f$ and $f$ are extracted from the posit decoding phase without modifications, the total exponent $te$ is introduced as $te= 2^{ES}\times k + e,$ in agreement with Equation \eqref{eqn:positRealValue}. Of course, $te$ can be both positive, zero or negative, while $e$ is a non-negative integer. We will use this intermediate representation as an helper, to compute the intermediate result of operations between two posits. Then we will normalize the intermediate result, hence we will transform FIR result into a proper posit. It is during this last phase that the posit rounding mechanism comes into play.

\subsection{Addition and Subtraction}

Addition and subtraction operations share several portion of arithmetic and logic parts. In general we want a FIR-represented posit $p_{out}$ such that: 
\begin{align*}\label{equ:addition_equation_001}
    \underbrace{\big[ (-1)^{s_1} \times 2^{te_1} \times (1.f_1) \big]}_{p_1} \pm \underbrace{\big[ (-1)^{s_2} \times 2^{te_2} \times (1.f_2) \big]}_{p_2} = \\ = \underbrace{ (-1)^{s_{out}} \times \overbrace{\big(2^{2^{ES}}\big)^{k_{out}} \times 2^{e_{out}}}^{2^{te_{out}}} \times (1.f_{out})}_{p_{out}}
\end{align*}
By supposing that $|p_1| \ge |p_2|$ we can transform the previous equation by extracting a scale factor $b = te_1 - te_2 \geq 0$ from the exponents, obtaining the following expression:
\[
p_{out}=(-1)^{s_1} \times 2^{te_1} \times \big[ \underbrace{(1.f_1) \pm (1.f_2) \times 2^{-b}}_{1.f_{out}}\big]
\]
After this step, we just need to normalize  $f_{sum} =(1.f_1) \pm (1.f_2) \times 2^{-b}$ to obtain the final $1.f_{out}$ value. With the \textit{addition} we need to ensure that the sum is strictly lower than 2, to be in a valid format. If $f_{sum}$ overflows 2, we scale it back by one position, summing $1$ to $te_1$. On the other hand, with the subtraction, we can normalize the result by subtracting the number of leading zeroes in $f_{sum}$  from $te_1$ and shifting the result of the same number of positions to the left. Note that, at the end of addition, when we shift right $f_{sum}$ of one position, we discard the last bit. If the discarded bit is $1$, we signal this outside, to preserve the information for final rounding phase.

\subsection{Multiplication}

As before, we have two posits: $p_1 = \left <s_1,k_1,e_1,f_1 \right >$ and $p_1 = \left <s_2,k_2,e_2,f_2 \right >.$ The goal is finding the FIR tuple $\left <s_{out}, te_{out}, f_{out}\right >$ such that:
\begin{align*}\label{equ:multiplication_equation_001}
    \big[ (-1)^{s_1} \times 2^{te_1} \times (1.f_1) \big] \times \big[ (-1)^{s_2} \times 2^{te_2} \times (1.f_2) \big] = \\ =(-1)^{s_{out}} \times \overbrace{\big(2^{2^{ES}}\big)^{k_{out}} \times 2^{e_{out}}}^{2^{te_{out}}} \times (1.f_{out})
\end{align*}
where, of course, $te_1= 2^{ES}\times k_1 + e_1$ and $te_2= 2^{ES}\times k_2 + e_2.$ 
Simplifying the equation, we obtain the following:
\begin{equation*}
    (-1)^{s_1 \oplus s_2}  \times \big[ 2^{te_1} \times 2^{te_2} \big] \times \big[ (1.f_1) \times (1.f_2) \big]
\end{equation*}
where $\oplus$ is the exclusive OR between the two sign bits $s_1$ and $s_2$.
As we can see, $te_{out} = te_1 + te_2$ while $1.f_{out}$ is the integer multiplication of the two fractions, readjusted for the normalization (along with the possible increment of $te_{out}$).

\subsection{Division}

Again, we have two posits $\left <s_{1,2},k_{1,2},e_{1,2},f_{1,2}  \right >$; the goal is finding the tuple $(s_{out}, te_{out}, f_{out})$  such that the following holds:\begin{equation}\label{equ:division_equation_001}
\footnotesize{
\frac{\big[ (-1)^{s_1} \times 2^{te_1} \times (1.f_1) \big]}{\big[ (-1)^{s_2} \times 2^{te_2} \times (1.f_2) \big]} = (-1)^{s_{out}} \times \overbrace{\big(2^{2^{ES}}\big)^{k_{out}} \times 2^{e_{out}}}^{2^{te_{out}}} \times (1.f_{out}) \,.
}
\end{equation}
Simplifying the left-hand side of (\ref{equ:division_equation_001}), we obtain
\begin{equation}
    \frac{(-1)^{s_1}}{(-1)^{s_2}} \times \frac{2^{te_1}}{2^{te_2}} \times \frac{(1.f_1)}{(1.f_2)}
\end{equation}
which similarly to the multiplication, suggests that $s_{out} = s_1 \oplus s_2$ and $te_{out} = te_1 - te_2$.
From the mathematical standpoint it comes off as not too dissimilar than a product.
If we consider that $(1.f)$ does not belong to $\mathbb{R}$, but instead it belongs to $ \mathbb{Q}$, we can multiply both numerator and denominator such that they are two integer numbers. Equation \eqref{divunsignedinteger00001} shows an example of this simplification, with the result being the one of an integer division.

\begin{equation}\label{divunsignedinteger00001}
\frac{1.\overbrace{011\dots 0001001}^{F \; bits}}{1.\underbrace{110\dots 0001111}_{F \; bits}} \equiv \frac{1011\dots 0001001 \times \cancel{2^{-F}}}{1110\dots 0001111 \times \cancel{2^{-F}}} \;.
\end{equation}

\subsection{Result normalization}

\begin{figure}
    \begin{center}
    \includegraphics[width=\linewidth]{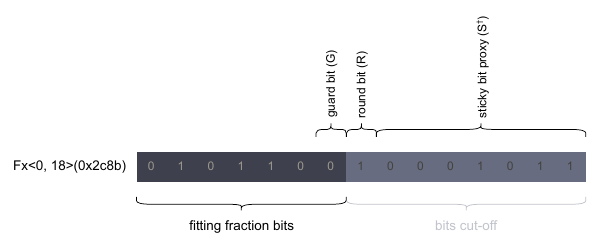}
    \caption{Bit-string layout of the fraction before rounding, with (G, R, S) bits highlighted.}
    \label{fig:fraction_before_rounding}
    \end{center}
\end{figure}

In the final stage we take the FIR output from the previous one and output a posit number (i.e. the result of the operation). The final posit $(s_{out},$ $k_{out}, e_{out}, f_{out})$ is computed from the resulting FIR $(s,te,f)$.
Firstly, we split $te$ into a posit regime $k'$ and exponent $e$:
\begin{equation}\label{k_and_exp_from_totalexp}
\begin{cases}
    k' \leftarrow \left \lfloor \dfrac{te}{2^{ES}} \right \rfloor \\
    k_{out} \leftarrow \text{clip}(k') \\
    e \leftarrow te - 2^{ES} \times k_{out} \,.
\end{cases}
\end{equation}
Note that $k' \neq k_{out}$ since it may result in a regime length higher than the maximum regime length for a posit$\left<N,x\right>$, that is $N-1$, including the stop bit. Typically this means that we need to \textit{clip} the value for $k'$ to the (maximum, or minimum) $k_{out}$ allowed. If the regime value is negative, its minimum value is $-(N-1)$; if the regime is positive, its maximum value is $N-2$, according to Eq. \eqref{eqn:regimeValue}. After we compute the number of regime bits, the actual size of the exponent is inferred from the regime field size and the posit parameters; the fractional field size is computed as the remaining bits, if any.
Finally, we need to accommodate for the fraction bits that fall off the posit size, if any. That turns out to be an implicit \textit{rounding to lowest}. Depending on the adopted rounding scheme, a few more operations must be considered. Assuming the  selected rounding scheme is \textit{round to nearest even} (the only one considered in the posit standard \cite{posit_standard_2022}), we consider 3 bits from the fraction  (Figure \ref{fig:fraction_before_rounding}):
\begin{itemize}
\item guard bit (G): the least significant bit of the sequence of digits that fit the fraction field of the final posit,
\item round bit (R): the most significant bit of the sequence of discarded digits,
\item sticky bit (S): the \textit{or-reduction}\footnote{adopting the Verilog nomenclature, the \textit{or-reduction} operator (\texttt{|}) applies the bitwise inclusive \textit{or} to the elements of a vector and returns a scalar.} of the sequence of bits to the right of the round bit (i.e. the discarded bits).
\end{itemize}
These bits will be used to complete the rounding of the posit, applying the \textit{round to nearest even} policy as described in \cite{posit_standard_2022}.

\section{Multi-stage Full Posit Processing Unit}\label{sec:fppu}

\begin{figure*}
    \centering
    \includegraphics[width=0.8\linewidth]
    {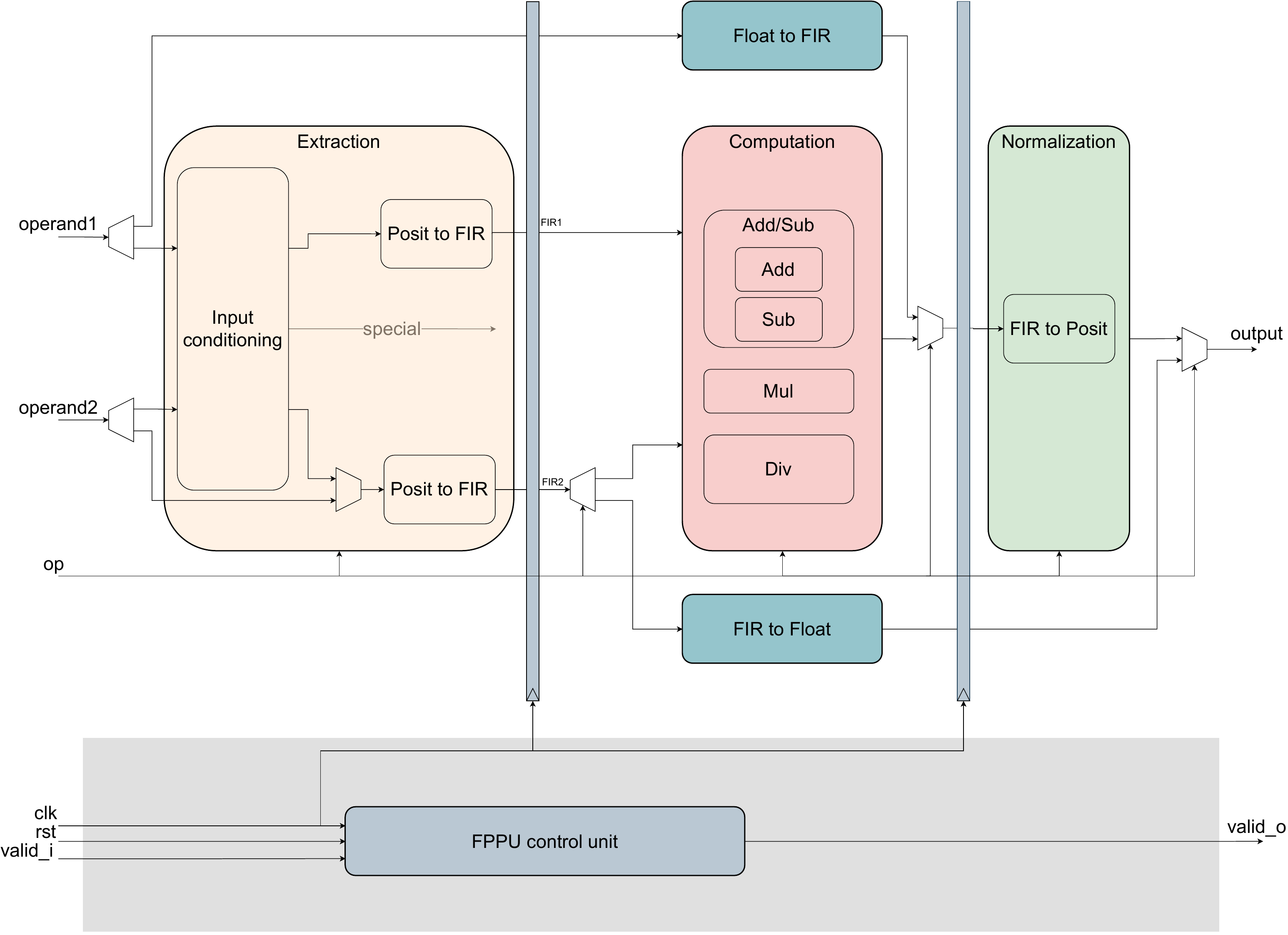} 
    \caption{3-stage Full PPU with control unit.}
    \label{fig:fppu_4s}
\end{figure*}

\begin{figure}
    \centering
    \includegraphics[width=\linewidth]{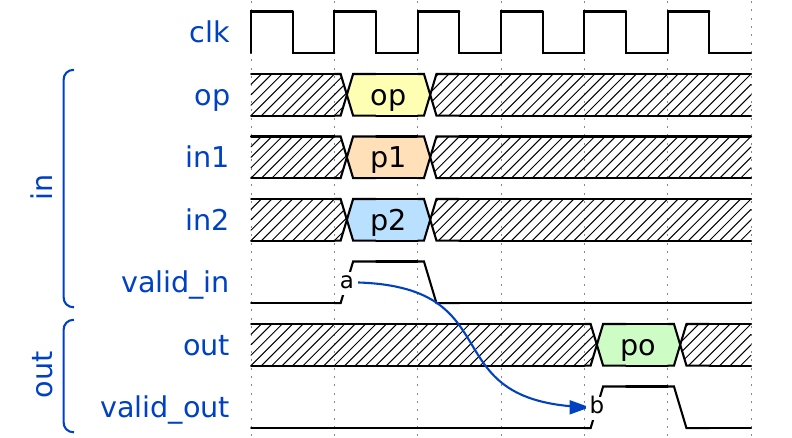}
    \caption{Example of interaction with the FPPU.}
    \label{fig:fppu_hs}
\end{figure}

This section presents the design of the FPPU. Similarly to previous section, we identify 3 main stages of execution inside the processing unit: i) decoding and input conditioning, ii) actual computation, iii) normalization and encoding of the result. To limit the length of pure combinatorial paths the FPPU has 4 pipelined stages corresponding to these 3 main stages of execution, with the computation phase split into two to take into account for the longer path in the division logic.
Figure \ref{fig:fppu_4s} shows a top-level schematic of the FPPU. The FPPU control unit has the task of signaling whether the current output is valid (i.e. 4 stages have been traversed by an instruction) or not. Figure \ref{fig:fppu_hs} shows an example of interaction with the FPPU with the submission of an operation identified by \texttt{OP} and the two operands \texttt{P1,P2}. When the operands are ready, \texttt{valid\_in} is set to active and after 3 cycles the FPPU produces a valid result \texttt{PO} signaled by \texttt{valid\_out}.

\subsection{The division algorithm}
This Section highlights some properties of the division algorithm implemented in the FPPU. In Section \ref{sec:positarith} we stated that the very last step in posit division is the integer division of the fractional parts. This operation is not as easy as multiplying or sum two integer numbers. Integer division is a problem that can be tackled in three main different ways \cite{muller_hardware_2010}:
\begin{itemize}
    \item Digit Recurrence algorithms: similar to pen and paper algorithm when dividing numbers. It produces a certain number of digits at each step by i) determining the next quotient digit(s), ii) multiplying such digits by the divider and iii) subtracting this result from the current reminder.
    \item Polynomial approximation: the division $\frac{x}{y}$ is computed multiplying $x$ with an approximation of $\frac{1}{y}$ using a polynomial expression.
    \item Iterative approximation: the reciprocate approximation is computed iteratively (e.g. Newton-Raphson).
\end{itemize}
Our architecture combines the last two families of algorithms to provide an accurate approximation of the reciprocate to be used in the subsequent multiplication.
In particular, we start from the idea of Chebyshev polynomials \cite{Hale2015} to provide an approximation of the reciprocate across the interval $(0.5,1)$. The expression for a third-order Chebyshev polynomial that approximate the $\frac{1}{x}$ function is: 
\begin{equation}\label{equ:3rd_order_Chebyshev_polynomial_equation}
f(x) = 5.65685 - 11.75737x + 10.64818x^2 - 3.54939 x^3 + \mathcal{O}(x^4) .
\end{equation}
The issue in using this approximation is that it requires a sequence of 6 fixed-point multiplications. An optimized approach is present in \cite{drom}, where the reciprocal approximation is computed as in Algorithm \ref{alg:drom_approx}, where $k_1,k_2$ are two parameter that we are going to evaluate to control the accuracy. This formulation has only 2 fixed point multiplication, since the $e \cdot 4$ can be implemented with a logical shift left of two positions.
\begin{algorithm}
\caption{Reciprocal approximation of $x$ from \cite{drom}.}\label{alg:drom_approx}
\begin{algorithmic}
\Require $x$, $k_1$, $k_2$
\Ensure $y = \frac{1}{x}$
\State $b \gets k_1 - x$
\State $c \gets x \cdot b$
\State $d \gets k_2 - c$
\State $e \gets d \cdot b$
\State $y \gets e \cdot 4$
\end{algorithmic}
\end{algorithm}
Expanding such routine in a polynomial we obtain an expression similar to \eqref{equ:3rd_order_Chebyshev_polynomial_equation}:
\begin{equation}
    f(x,k_1,k_2) = 4 k_1 k_2 - 4(k_1^2 + k_2) x + 8 k_1 x^2 - 4 x^3 .
\end{equation}
Instead of using the $k_1,k_2$ values proposed in \cite{drom} we set up a minimization problem of the error function $e^2$ of the reciprocal approximation. In particular we defined this function as:
\begin{equation}
    e^2(k_1, k_2) =  \int_{1/2}^{1} rerr^2(x, k_1, k_2)\ dx \, ,
\end{equation}
where $rerr$ is the relative error between the approximated and the exact inverse function. We then use this function to solve the following problem:
\begin{equation}\label{equ:opt_k1_k2_eq}
(k_{1_{opt}}, k_{2_{opt}}): \min\{e^2(k_1, k_2)\} .
\end{equation} 
The obtained solution gives $k_{1_{opt}} = 1.4567844114901045$ and  $k_{2_{opt}} = 1.0009290026616422$, yielding a $36.4\%$ improvement over \cite{drom}. We can pair this solution with a round of Newton Raphson using the output of this stage as starting condition to refine the accuracy of the approximation.


We compared the accuracy of our solution to the solution of PACoGen \cite{PACoGen}, that employs a pre-computed look-up table for reciprocal approximation and as input of a round of Newton-Raphson. Table \ref{table:comparison_div_against_pacogen_table} reports this comparison. In the table, the \textbf{IN} column is the number of fraction bits from the posits used to index the LUT, while \textbf{OUT} is the number of bits of the reciprocal approximation of the fraction; \textbf{NR} indicates the number of Newton-Raphson rounds used and \textbf{wrong [\%]} states the error percentage of the division results when compared to a software golden model for posit computation.

\begin{table}
\begin{center}
\caption{Percentages of posits $P\langle N,ES\rangle$ inexact division results. PACoGen version \cite{PACoGen} vs proposed.}
\begin{tabular}{ccccccccc}
    \toprule
     & & \multicolumn{4}{c}{PACoGen} & & \multicolumn{2}{c}{proposed} \\
    \cmidrule{3-6} \cmidrule{8-9}
    N & ES & IN & OUT & NR & wrong [\%] & & NR & wrong [\%] \\
    \midrule \midrule
    8 & 0 & 8 & 9 & 0 & 4.8 & & 1 & \textbf{1.4} \\
    8 & 1 & 8 & 9 & 0 & 5.4 & & 1 & \textbf{1.2} \\
    8 & 2 & 8 & 9 & 0 & 9.3 & & 1 & \textbf{2.1} \\
    8 & 3 & 8 & 9 & 0 & 13.5 & & 1 & \textbf{4.2} \\
    8 & 4 & 8 & 9 & 0 & 16.4 & & 1 & \textbf{7.5} \\
    \midrule
    16 & 0 & 8 & 9 & 1 & 10.0 & & 1 & \textbf{1.5} \\
    16 & 1 & 8 & 9 & 1 & 10.0 & & 1 & \textbf{0.6} \\
    16 & 2 & 8 & 9 & 1 & 8.8 & & 1 & \textbf{0.5} \\
    16 & 3 & 8 & 9 & 1 & 9.0 & & 1 & \textbf{0.1} \\
    \bottomrule
\end{tabular}
\label{table:comparison_div_against_pacogen_table}
\end{center}
\end{table}

\section{RISC-V ISA Extension and compiler support}\label{sec:rvisaposit}

This section briefly presents the RISC-V ISA extension for Posits and the related compiler support. We decided to reuse the existent RISC-V registers (i.e. \texttt{x1...x31}). Since posits can be treated as signed integers inside the architecture, we followed the RV32I base integer instruction set from the RISC-V standard for arithmetic operations. In particular, \text{ADD, SUB, MUL, DIV} posit operations are encoded as \textit{R-type} instructions (see \cite{riscvisa}). We leveraged the custom opcode space \texttt{0x0B} from the RISC-V standard to add the new instructions to the ISA. In Table \ref{sec:rvisaposit} we reported the listing of posit instructions used for the RISC-V ISA extension proposed in this work. Besides the posit arithmetic operations we also added conversion instructions between posits and binary32 numbers, so that we can enable the use of binary32 numbers as frontend while maintaining posit computation as backend in HW. To enable the use of the novel instructions in SW, we provided a set of intrinsic functions to map high-level C/C++ function call to the underlying machine code. Doing this, we do not need to change the RISC-V compiler but we generate the correct assembly for the posit ISA at compile time. 

\begin{table}
\caption{Instruction listing for posit arithmetic ISA extension}
\begin{small}
\begin{center}
\begin{tabular}{cccccccccccl}
\cline{2-11}
&
\multicolumn{4}{|c|}{funct7} &
\multicolumn{2}{c|}{rs2} &
\multicolumn{1}{c|}{rs1} &
\multicolumn{1}{c|}{funct3} &
\multicolumn{1}{c|}{rd} &
\multicolumn{1}{c|}{opcode} & R-TYPE \\
\cline{2-11} \\
\cline{2-11}

&
\multicolumn{4}{|c|}{1100000} &
\multicolumn{2}{c|}{rs2} &
\multicolumn{1}{c|}{rs1} &
\multicolumn{1}{c|}{000} &
\multicolumn{1}{c|}{rd} &
\multicolumn{1}{c|}{0001011} & PADD \\
\cline{2-11}

&

\multicolumn{4}{|c|}{1101010} &
\multicolumn{2}{c|}{rs2} &
\multicolumn{1}{c|}{rs1} &
\multicolumn{1}{c|}{001} &
\multicolumn{1}{c|}{rd} &
\multicolumn{1}{c|}{0001011} & PSUB \\
\cline{2-11}

&
\multicolumn{4}{|c|}{1100000} &
\multicolumn{2}{c|}{rs2} &
\multicolumn{1}{c|}{rs1} &
\multicolumn{1}{c|}{010} &
\multicolumn{1}{c|}{rd} &
\multicolumn{1}{c|}{0001011} & PMUL \\
\cline{2-11}

&
\multicolumn{4}{|c|}{1100000} &
\multicolumn{2}{c|}{rs2} &
\multicolumn{1}{c|}{rs1} &
\multicolumn{1}{c|}{100} &
\multicolumn{1}{c|}{rd} &
\multicolumn{1}{c|}{0001011} & PDIV \\
\cline{2-11}

&
\multicolumn{4}{|c|}{rs3 | 00} &
\multicolumn{2}{c|}{rs2} &
\multicolumn{1}{c|}{rs1} &
\multicolumn{1}{c|}{000} &
\multicolumn{1}{c|}{rd} &
\multicolumn{1}{c|}{0101011} & PFMADD \\
\cline{2-11}

\end{tabular}
\end{center}
\end{small}
\label{tab:rvxposit}
\end{table}

We report an example of intrinsic and relative call from C code in Listing \ref{lst:padd_snippet}. The \texttt{register} keyword suggests the compiler to put the values of the input operands and the result in three registers, since the instruction type is R-type. At lines \texttt{6,7,8} we set up the \texttt{opcode, funct3, funct7} parameters for the specific operation (this is the only part that varies between different operations).

\begin{lstlisting}[language=Cisa, caption={Example of intrinsic for a posit addition operation.}, captionpos=b, label={lst:padd_snippet}]
posit_t padd(long a, long b) {
  register int p1 asm("a1") = a;
  register int p2 asm("a2") = b;
  register int result asm("a0");
  __asm__(
  ".set op,0xb\n"
  ".set opf1,0\n"
  ".set opf2,0x6A\n"
  ".byte  op|((r%[result]&1) <<7),"  
  "((r%[result]>>1)&0xF)|
  (opf1<<4)|((r%1 &1) <<7),"  
  "((r%2&0xF) << 4) | ((r%1>>1)&0xF),"  
  "((r%2>>4)&0x1)|(opf2 << 1)\n"
  : [result] "=r"(result)
  : "r"(p1), "r"(p2), "[result]"(result));
  return result;
}
\end{lstlisting}

A more complex example is shown in Listing \ref{lst:gemm_snippet} and \ref{lst:conv_snippet} with the implementation of a simple square matrix multiplication and a $3\times 3$ convolution.

\begin{lstlisting}[language=Cisa2, caption={Square matrix-matrix multiplication using posit intrinsics.}, captionpos=b, label={lst:gemm_snippet}]
void gemm(posit_t *a, 
          posit_t *b,
          posit_t *c, 
          int n) {
 for (int i = 0; i < n; i++) {
  for (int j = 0; j < n; j++) {
   posit_t sum = 0;
   for (int k = 0; k < n; k++) {
    sum = padd(sum, pmul(a[i*n+k], b[k*n+j]));
   }
   c[i * n + j] = sum;
  }
 }
}
\end{lstlisting}

\begin{lstlisting}[language=Cisa2, caption={3$\times$3 convolution using posit intrinsics.}, captionpos=b, label={lst:conv_snippet}]
void conv3x3(posit_t *a, 
             posit_t *f, 
             posit_t *c,
             int n) {
 for (int i = 0; i < n; i++) {
  for (int j = 0; j < n; j++) {
   posit_t sum = 0;
   for (int k = 0; k < 3; k++) {
    for (int l = 0; l < 3; l++) {
     sum = padd(sum, pmul(a[(i+k)*n+j+l],f[k*3+l]));
    }
   }
   c[i * n + j] = sum;
  }
 }
}
\end{lstlisting}

We cover more examples and tests in Section \ref{sec:fppuintegr} with the verification and validation of the FPPU inside the Ibex Core.

\section{The Ibex core and FPPU integration}\label{sec:fppuintegr}

The Ibex Core \cite{schiavone_zeroriscy} is a small 32-bit RISC-V core with a 2-stage pipeline. It supports the Integer (I) or Embedded (E), Integer Multiplication and Division (M), Compressed (C), and B (Bit Manipulation) extensions. Since it does not employ a floating point unit, it is particularly useful to test the impact of adding real number arithmetic using the FPPU. 

We put the FPPU alongside the Arithmetic Logic Unit (ALU) inside the execution stage of the Ibex pipeline. Since we did not add specific registers for the posit operations, there was no need to modify the \textit{register file} but only the \textit{decoder} module by adding the instructions designed in Section \ref{sec:rvisaposit}. 
\begin{figure}
    \centering
    \includegraphics[width=0.7\linewidth]{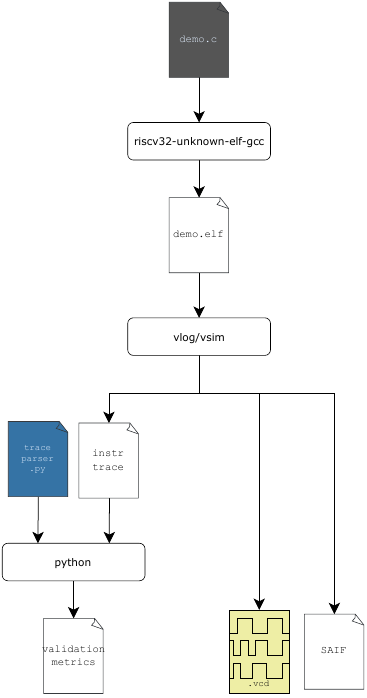}
    \caption{Workflow diagram for compiling, simulating and validating the instruction set added to the RISC-V ISA.}
    \label{fig:wflow}
\end{figure}

After the integration, we tested and validated the integration exploiting the RTL simulation with binaries compiled for the \texttt{RV32IM} architecture as shown in Fig. \ref{fig:wflow}. Using the instruction tracer inside the Ibex core we managed to dump all the instructions executed by the core during the RTL simulation, including the newly added posit instructions. We fed the output of the tracer to a trace parser (written in python language) to evaluate the output of each posit instruction, comparing it to the same software golden model used for validating the standalone FPPU. Furthermore, we used the trace parser to assess the accuracy of each operation, comparing the result to the IEEE binary32 correspondent operation result. 

\subsection{Integration tests and validation}

To test the compliance of the FPPU HDL design vs. a posit golden model, we tested the overall system with different DNN kernels on $32\times32$ matrices (i.e. size of images for MNIST/CIFAR10 datasets). In detail, we tested a matrix-matrix multiplication, a $3\times3$ convolution and a $4\times4$ average pooling. Each test was run on the 8-bit FPPU and on the 16-bit FPPU. We collected the instruction traces and fed them to the trace parser we described before. We then collected two sets of results: i) the accuracy w.r.t the posit golden model and ii) the normalized mean error w.r.t the same operation executed with binary32 format. Let $r_p^i, r_f^i$ be, the i-th result for, respectively, the posit and the binary32 operation, we computed the normalized mean error $\overline{e_{op}}$  for a given operation $op$ as follows:
\[
\overline{e_{op}} = \frac{1}{N}\sum_{i}^{N} \left | \frac{r_p^{i} - r_f^{i}}{r_f^i} \right | .
\]

Furthermore, we report more complex accuracy tests done with the posit format with the following benchmarks:
\begin{itemize}
    \item Small datasets (MNIST, GTSRB and CIFAR-10) on LeNet5 convolutional neural network \cite{lecunlenet,lecun-mnisthandwrittendigit-2010,stallkamp2011gtrsb}
    \item Big datasets (ImagenetV2 \cite{recht2019imagenet}, VOC2007 \cite{pascal-voc-2007}) on complex DNN models (EfficientNetB0 \cite{tan2020efficientdet} and Single Shot Detector SSD300).
\end{itemize}

\begin{figure}
    \centering
    \includegraphics[width=\linewidth]{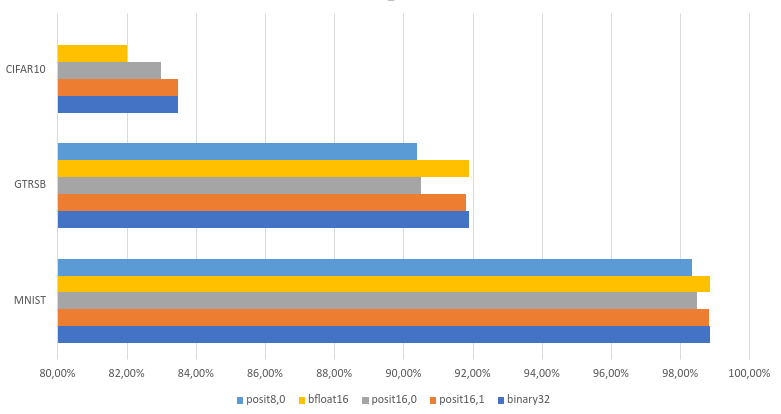}
    \caption{Accuracy comparison between 8-bit, 16-bit posit formats and 32-bit IEEE binary32 on the LeNet-5 convolutional neural network.}
    \label{fig:small_nns}
\end{figure}

\begin{figure}
    \centering
    \includegraphics[width=\linewidth]{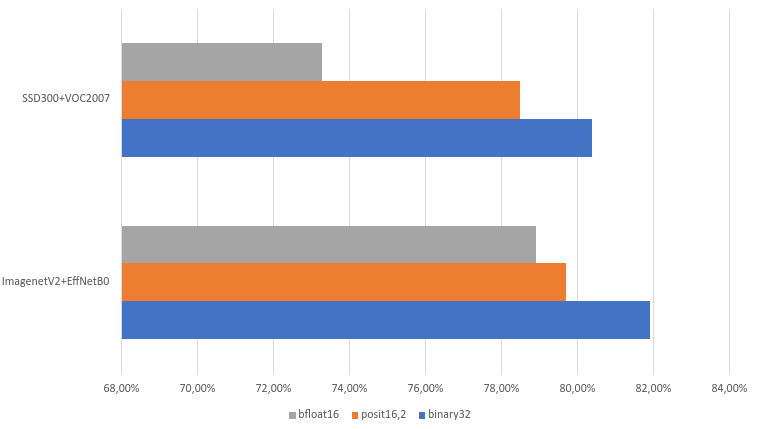}
    \caption{Accuracy comparison between 16-bit formats (posit, bfloat16) and 32-bit IEEE binary32 on complex DNN task.}
    \label{fig:big_nns}
\end{figure}

Table \ref{tab:binary32_error} shows the normalized mean error of posit operations compared to the correspondent binary32 ones. Figures \ref{fig:small_nns} and \ref{fig:big_nns} show accuracy performance of posit employed in several neural network tasks.

\begin{table}[]
\caption{Normalized mean error of FPPU operations vs. correspondent binary32 operation in several linear algebra tasks.}
\label{tab:binary32_error}
\begin{tabular}{c|cc|cc|cc|}
\cline{2-7}
      & \multicolumn{2}{c|}{Conv (3x3 filter)}       & \multicolumn{2}{c|}{GEMM}           & \multicolumn{2}{c|}{Average Pooling 4x4} \\ \cline{2-7} 
      & \multicolumn{1}{c|}{p$\left < 8,0 \right >$}  & p$\left < 16,2 \right >$  & \multicolumn{1}{c|}{p$\left < 8,0 \right >$}  & p$\left < 16,2 \right >$  & \multicolumn{1}{c|}{p$\left < 8,0 \right >$}     & p$\left < 16,2 \right >$   \\ \cline{1-7} 
 p.mul & \multicolumn{1}{c|}{0.042} & 0.004  & \multicolumn{1}{c|}{0.019} & 0.003  & \multicolumn{1}{c|}{-}        & -        \\
p.add & \multicolumn{1}{c|}{0.025} & 0.0004 & \multicolumn{1}{c|}{0.016} & 0.0007 & \multicolumn{1}{c|}{0.019}    & 0.0002   \\
p.div & \multicolumn{1}{c|}{-}     & -      & \multicolumn{1}{c|}{-}     & -      & \multicolumn{1}{c|}{0.002}    & 0        \\ \hline
\end{tabular}
\end{table}

\section{FPPU and Ibex characterization}\label{sec:ppuchar}

\begin{figure*}
    \centering
    \includegraphics[width=\linewidth]{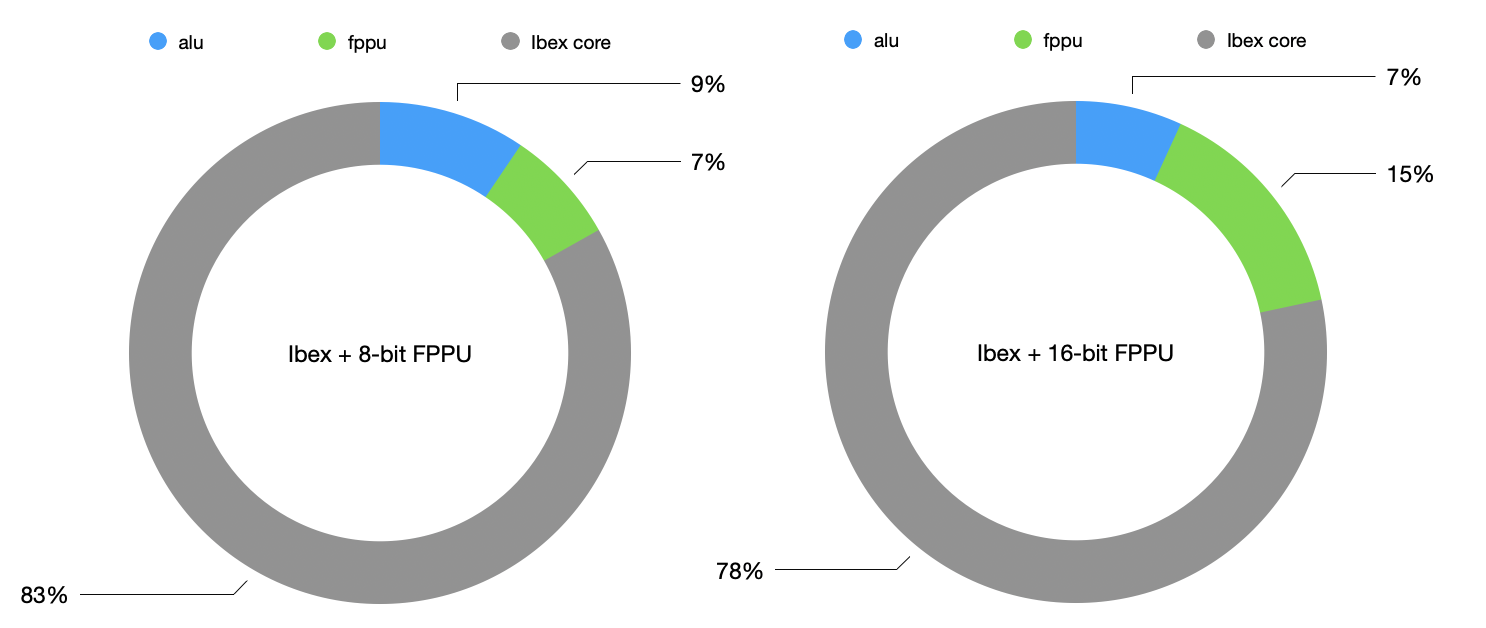}
    \caption{Percent Area utilization (LUTs) of the FPPU with $\text{Posit}\left < 8,2 \right >$ (left) and $\text{Posit}\left < 16,2 \right >$ (right) and the other components of Ibex.}
    \label{fig:percent_ibex_luts}
\end{figure*}

\begin{figure}
    \centering
    \includegraphics[width=\linewidth]{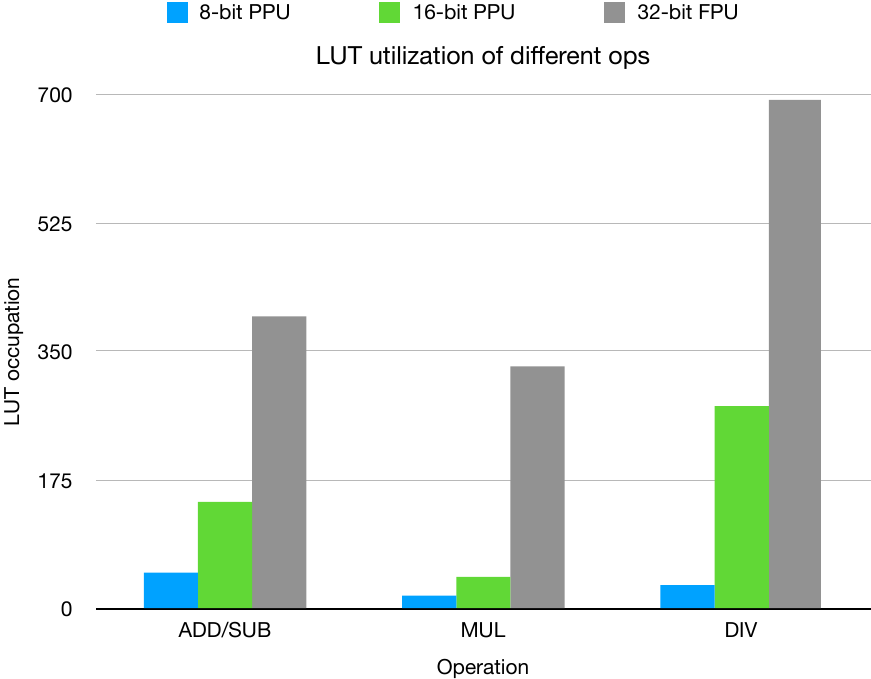}
    \caption{Comparison between absolute area utilization (LUTs) of 8-bit, 16-bit FPPU and 32-bit FPU operations.}
    \label{fig:abs_ops_luts}
\end{figure}

In this section we characterize the FPPU by implementing it, alone or integrated within the Ibex RISC-V core, targeting Xilinx Alveo U280 (\texttt{xcu280-fsvh2892-2L-e}). From the synthesis process we extracted area and power metrics. Furthermore, we instantiated the Pulpino SoC \cite{7864441} with the CV32E40P core \cite{schiavone_zeroriscy}, that embeds a 32-bit floating point unit, in order to compare the area occupation of the operation implementations in both the solutions.
Fig. \ref{fig:percent_ibex_luts} shows the area occupation of the arithmetic logic unit (ALU) and the FPPU relatively to the total area of the Ibex core. As we can see, with an 8-bit PPU we are able to provide real number arithmetic capabilities to the core with an area cost that is less than the cost of the original Ibex ALU. Fig. \ref{fig:abs_ops_luts} shows the comparison between FPPUs and FPU in terms of area occupation of the logic related to ADD, MUL and DIV operations. As we can see, using half of the bits for the representation results in a area cost that is less then half of the area for the 32-bit counterpart. The area cost for the comparison is not reported, since posits can be compared as signed integers, while binary32 numbers require dedicated circuits.  
We exploited the SAIF file produced at the end of the workflow shown in Fig.  \ref{fig:wflow} to estimate the dynamic power of the FPPU component (see also \cite{parma2023piccoli}) in the four different operations. Table \ref{tab:fppu_power_detail} reports the details for the four arithmetic operations.
The results refer to Alveo FPGA implementation, and the dynamic power at a clock frequency of 20 MHz (with a maximum achievable frequency of 100 MHz)  is below one mW for each 8-bit posit operation. The maximum FPPU peak throughput at 100 MHz is 33 MOps/s considering a 30 ns latency for the FPPU component with three pipeline stages. 

\subsection{SIMD configuration}

Considering that we are under-using the 32-bit registers when employing 8-bit or 16-bit posits, we can think of increasing the number of FPPUs that elaborate in parallel, similarly to a Single Instruction Multiple Data (SIMD) paradigm. This can be done transparently to the instruction caller and with the same opcode. When using only one FPPU we just need to put the posit arguments in 8 (or 16) least significant bits. If we want to compute more posit operations in parallel, we would just need to put another three posits in the remaining 24 bits of the register (or another one posit in the remaining 16 bits). This can be easily implemented by reproducing the same FPPU 2 or 4 times (respectively, for 16-bit posits or for 8-bit posits). Then we can feed the same operand, valid, reset and clock inputs to all the FPPUs, while the two operands are constituted by portions of the source registers. The output is then concatenated from all the FPPU outputs into the destination register. We have verified that this approach would increase the throughput of the FPPU to 132 MOps/s for 8-bit posits and to 66 MOps/s for 16-bit posits. Of course the main difficulties arise from the software side, where we need to change the C++ source code (or any other programming language source code) to take advantage from this extended feature, and this can be challenging for complex applications. Indeed, we need either a compiler supporting automatic vectorization of the code  or a partial rewrite of mathematical kernels to explicitly support the SIMD paradigm.

\begin{table}[]
\caption{Detail of mean and standard deviation of dynamic power measurement of the FPPU component.}
\label{tab:fppu_power_detail}
\centering
\begin{tabular}{l|c|c|}
\cline{2-3}
    & 8-bit FPPU (mW) & 16-bit FPPU (mW) \\ \hline
 ADD &  $<1$         & $1$           \\
 SUB &  $<1$         & $1$           \\
 MUL &  $<1$         & $1$           \\
 DIV &  $1$          & $2$           \\ \hline
\end{tabular}
\end{table}

\section{Discussions and Conclusions}\label{sec:concl}
In this work we presented the design and integration of a full posit processing unit inside the low-power Ibex RISC-V core extending the RISC-V ISA to support posit arithmetic operations (add/sub, mul, div, posit-to-float/float-to-posit and FMA) as well as vectorized posit operations in a SIMD configuration of 4 units for 8-bit posits and 2 units for 16-bit posits. We provided several integration and validation test against both a posit golden model and a series of baseline binary32 neural network models (LeNet-5, EfficientNet, SSD300), showing at most a degradation of $4\%$ in terms of inference accuracy on the object detection tasks (PascalVOC), while obtaining negligible drops in accuracy for 16-bit posits in the other simpler tasks (MNIST, CIFAR10, GTSRB).
With respect to other works in literature like PaCoGen \cite{PACoGen}, this work allows for an improved division algorithm in terms of accuracy. We characterized the unit on power and area comparing in particular the occupation of the unit in relation to the the full-core, showing that the posit-8 FPPU complexity is even lower than the Ibex ALU. The increase in area occupation of the FPPU is 7\% for 8-bit posits and 15\% for 16-bit posits. 
When compared to the FPU used by the CV32E40P RISC-V core \cite{schiavone_zeroriscy}, the FPPU area occupation is less than half for 16-bit posits (for the same accuracy of binary-32) and 1 order of magnitude lower for 8-bit posits (for an accuracy loss within $4\%$).
When compared to work in \cite{ppulight} it has to be noted that the Light PPU gives to a RISC-V core the support to Posit with a limited complexity overhead and no speed overhead, but the support is limited to float-to/from-posit conversion, Hence, \cite{ppulight} is not effective for Posit computation since it requires that posit are converted in float, then processed by the FPU and then the results converted back in posit. Instead, the FPPU gives direct HW support to posit operations (sub, add, mul, div, inversion, conversion) so that computation in posit domain becomes efficient vs floating-point representation. Since the FPPU replaces in RISC-V the FPU the complexity overhead of the FPPU is lower than the overhead of Light PPU plus the FPU. 

\section*{Acknowledgments}
Work partially supported by H2020 projects EPI-SGA2 (grant no. 101036168 ) and TEXTAROSSA (grant no. 956831,  \url{https://textarossa.eu/} ). In addition we wish to thank the Italian Ministry of Education and Research (MIUR) for funding this research within the framework of the FoReLab project (Departments of Excellence).

\bibliographystyle{IEEEtran}
\bibliography{Bibliography}

\begin{IEEEbiography}[{\includegraphics[width=1in,height=1.25in,clip,keepaspectratio]{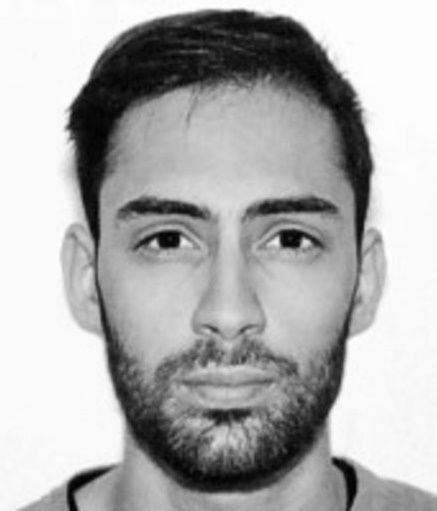}}]{Federico Rossi} is a PhD student of the Information Engineering Department at University of Pisa. In 2019 he received his Master Degree in Computer Engineering \textit{magna cum laude}. He is currently involved in the European Processor Initiative (EPI2), the TextaRossa and EuPilot european projects. His research topics include alternative real  number representations and their applications to Deep Neural Networks for the automotive environment.
\end{IEEEbiography}

\begin{IEEEbiography}[{\includegraphics[width=1in,height=1.25in,clip,keepaspectratio]{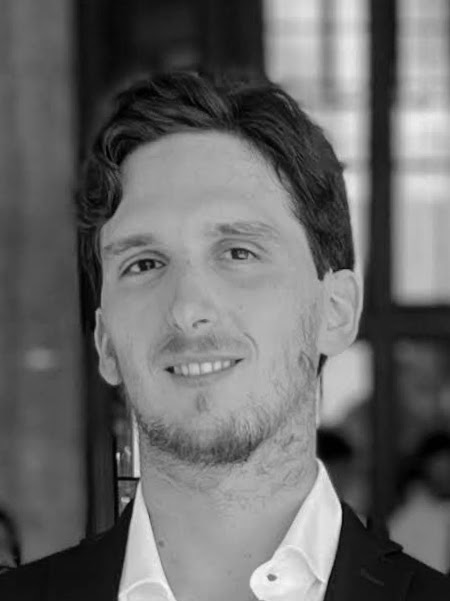}}]{Francesco Urbani} is a research fellow at the Department of Information Engineering at the University of Pisa. He received his Master's Degree in Electronic Engineering in 2022. He is currently involved in the European Processor Initiative (EPI) SGA-2 project. His research topics include alternative number representations for AI applications on the edge. 
\end{IEEEbiography}

\begin{IEEEbiography}[{\includegraphics[width=1in,height=1.25in,clip,keepaspectratio]{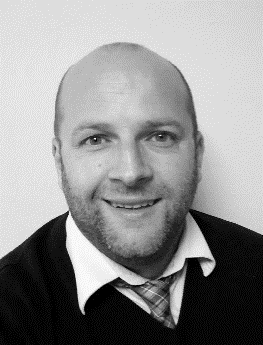}}]{Marco Cococcioni} (SM’12) received   the   Laurea   degree   in   2000   and   the Diploma degree in 2001 in Computer Engineering from  University  of  Pisa  and  Scuola  Superiore S. Anna,   respectively,   both   with   magna   cum laude.  In  2004  he  earned  the  Ph.D.  degree  in Computer  Engineering  at  the  University  of  Pisa. After working as a post-doc in the same department, in 2010-2011 he spent two years as Senior Visiting Scientist at the NATO Undersea Research Centre (now CMRE) in La Spezia, Italy. For his collaboration with CMRE he obtained the NATO Scientific Achievement Award in 2014. Since 2016 he is an Associate Professor at the Department of Information Engineering of the University of Pisa. He is in the editorial board of several journals indexed by Scopus. He is member of three IEEE task forces: Genetic Fuzzy Systems, Computational Intelligence in Security and Defense, and Intelligent System Application.  Prof. Cococcioni has co-authored more than 100 contributions to international journals and conferences and he is a Senior Member of both IEEE and ACM (Association for Computing Machinery). He has been involved in five H2020 European Projects (EPI SGA-1, EPI SGA-2, TextaRossa, EuPilot, and EUPEX).
\end{IEEEbiography}

\begin{IEEEbiography}[{\includegraphics[width=1in,height=1.25in,clip,keepaspectratio]{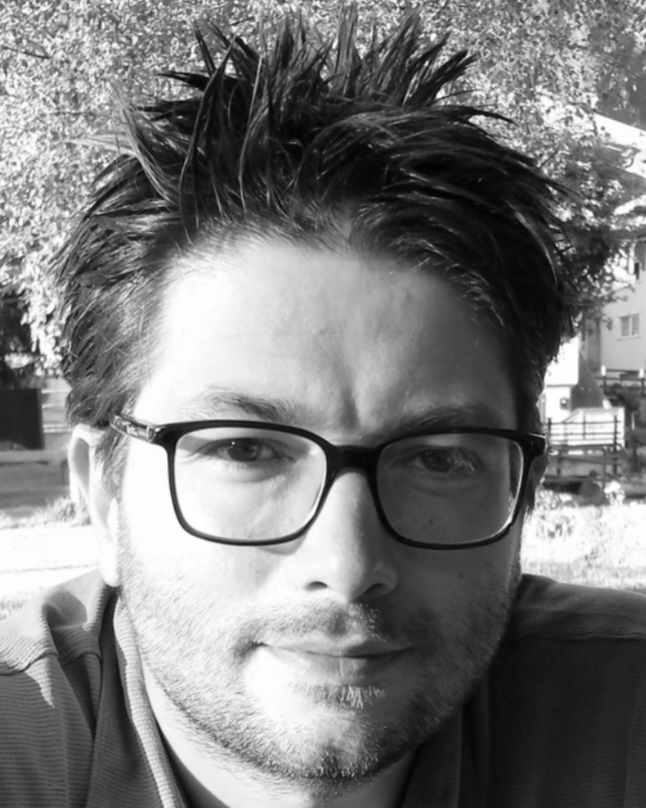}}]{Emanuele Ruffaldi} (SM’18) is senior software engineer at Medical Microinstruments Inc. working on robotic assisted microsurgery. Formerly he has been Assistant Professor at Scuola Superiore Sant’Anna in the Perceptual Robotics laboratory, Pisa, Italy. His research interests are in the field of machine learning for HRI and embedded artificial intelligence. He is Senior IEEE Member and has served IEEE as Publicity Chair for the Haptics TC.
\end{IEEEbiography}

\begin{IEEEbiography}[{\includegraphics[width=1in,height=1.25in,clip,keepaspectratio]{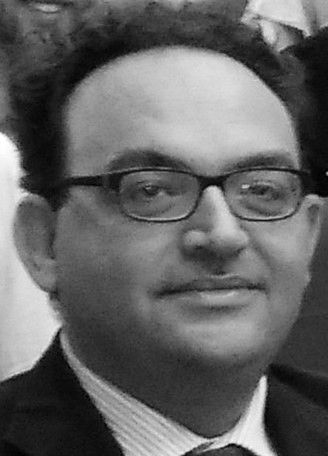}}]{Sergio Saponara} (SM’13) is Full Professor of Electronics at University of Pisa, where he got Master degree cum laude and Ph.D. degree. In 2012 he was a Marie Curie Research Fellow in IMEC. He is an IEEE Distinguished Lecturer and co-founder of special interest group on IoT of both IEEE CAS and SP societies. He is the director of I-CAS lab, of Crosslab Industrial IoT, of the Summer School Enabling Technologies for IoT. He is associate editor of several IEEE and Springer Journals. He co-authored more than 300 scientific publications and 18 patents. He is the leader of many funded projects by EU and by companies like Intel, Magneti Marelli, Ericsson and PPC.
\end{IEEEbiography}

\end{document}